\documentclass[twocolumn,english,showpacs,showkeys]{revtex4}
\usepackage[T1]{fontenc}
\usepackage[latin9]{inputenc}
\usepackage{array}
\usepackage{graphicx}
\usepackage{setspace}
\usepackage{babel}
\usepackage{amsmath}

\providecommand{\tabularnewline}{\\}

\begin{document}

\title{Magnetic properties of exchange biased and of unbiased oxide/permalloy thin layers: a ferromagnetic resonance and 
Brillouin scattering study}

\author{Fatih Zighem, Yves Roussigné, Salim Mourad Chérif and Philippe Moch}
\affiliation{\textit{Laboratoire des Propriétés Mécaniques et Thermodynamiques des Matériaux,
CNRS UPR 9001, Université Paris Nord, 93430 Villetaneuse, France}}

\author{Jamal Ben Youssef}
\affiliation{\textit{Laboratoire de Magnétisme de Bretagne, CNRS/FRE 3117, Université de Bretagne Occidentale,
29285 Brest, France}}

\author{Fabien Paumier}
\affiliation{\textit{Institut Pprime, CNRS UPR 3346, Université de Poitiers, SP2MI-BP 30179, 
86962 Chasseneuil-Futuroscope, France}}


\begin{abstract}
Microstrip ferromagnetic resonance and Brillouin scattering are used
to provide a comparative determination of the magnetic parameters
of thin permalloy layers interfaced with a non-magnetic (Al$_{2}$O$_{3}$)
or with an antiferromagnetic oxide (NiO). It is shown that the perpendicular
anisotropy is monitored by an interfacial surface energy term which
is practically independent of the nature of the interface. In the
investigated interval of thicknesses (5-25 nm) the saturation magnetisation
does not significantly differ from the reported one in bulk permalloy.
In-plane uniaxial anisotropy and exchange-bias anisotropy are also
derived from this study of the dynamic magnetic excitations and compared
to our independent evaluations using conventional magnetometry
\end{abstract}
\maketitle

\section{Introduction}
In magnetically ordered solids, the magnetic properties near a surface
or an interface may differ in many respects from the observed one
inside the bulk \cite{neel,bruno,roussigne95}. These differences are attributed to
the reduced symmetry, to the lower coordination number and to the
availability and role of highly localised surface and interface states
inducing modified magnetic structures, sources of interesting magnetic
behaviours. These phenomena which are generally localised within a
few atomic layers are phenomenologically treated by means of surface
anisotropies which provoke spin rearrangements inside thin magnetic
films \cite{bruno}. This paper focuses on metallic permalloy/oxide interfaces,
comparing the cases of a nonmagnetic oxide (Al$_{2}$O$_{3}$) with
the case of an antiferromagnetic one (NiO). In this last situation,
various interfacial effects have been put in evidence previously,
such as hysteresis loop shifts and increased coercivity \cite{meiklejohn}, training
effects \cite{meiklejohn,nogues} and rotatable anisotropy \cite{mcmichael}. In particular,
the so called exchange bias field was discovered nearly 50 years
ago \cite{meiklejohn} and has given rise to a large number of experimental and
theoretical publications \cite{nogues,binek,mcmichael,nogues2005,berkowitz1999,stamps2000}.
Up to now, most of the experimental studies are based on hysteretic measurements and interpreted in terms
of exchange bias ($H_j$) and coercive ($H_c$) fields. In contrast, only
few ferromagnetic resonance (FMR) and Brillouin light scattering (BLS)
studies, analysing the dynamic magnetic properties of exchange biased
bilayers, are available  \cite{mcmichael,stamps2000,stoecklein,geshev,blachowicz,wee,rezende1,zighem1}. The present paper takes
advantage of both techniques (more explicitely; retro-BLS and microstrip(MS)-FMR
in view of interpreting this dynamics in interfaced polycrystalline
Ni$_{81}$Fe$_{19}$ thin films of various thicknesses. In addition, a careful
examination of the interfaces was performed through high resolution
electronic transmission microscopy (HRTEM) using a JEOL 3010 microscope
(300 kV, LaB6, 0.19 nm point resolution).

The paper is organized as follows: in Section 2, we briefly describe
the model used for the interpretation of the experimental data. Section
3 presents the two above cited techniques: BLS and MS-FMR. The experimental
results are analysed and discussed in Section 4.

\section{Theoretical background}

For an uncoupled permalloy layer, the volume magnetic energy density
is written as: \begin{eqnarray}
\varepsilon_{0}=\varepsilon_{zee.}+\varepsilon_{dip.}+\varepsilon_{exch.}+\varepsilon_{anis.}\label{eq:1}\end{eqnarray}

where the first three terms stand for the Zeeman, the dipolar and
the exchange contributions, respectively. The last term represents
the anisotropy contribution and can be expressed as \cite{neel,bruno,roussigne95}:

\begin{eqnarray}
\varepsilon_{anis.}=-\frac{(\vec{M}\cdot\vec{n})^{2}}{M^{2}}K_{\perp}-\frac{(\vec{M}\cdot\vec{u})^{2}}{M^{2}}K_{u}
\label{eq:2}
\end{eqnarray}

In equation (2), $\vec{M}$ is the magnetisation. The unit vector
$\vec{n}$ is normal to the film and consequently, $K_{\perp}$ represents
a uniaxial perpendicular anisotropy parameter. In addition, $\vec{u}$
is an in-plane unit vector and $K_{u}$ stands for an in-plane anisotropy
parameter. In most cases, $K_{\perp}$ consists in an effective uniaxial anisotropy
parameter which results from the addition of a bulk (presumably magnetocrystalline)
term and of a surface term which depends on the film thickness $t$:
\begin{eqnarray}
K_{\perp}=K_{\perp B}+\frac{K_{\perp S}}{t}\label{eq:3}
\end{eqnarray}

\begin{figure}
\begin{center}
\includegraphics[bb=30bp 250bp 375bp 580bp,clip,width=8.5cm]{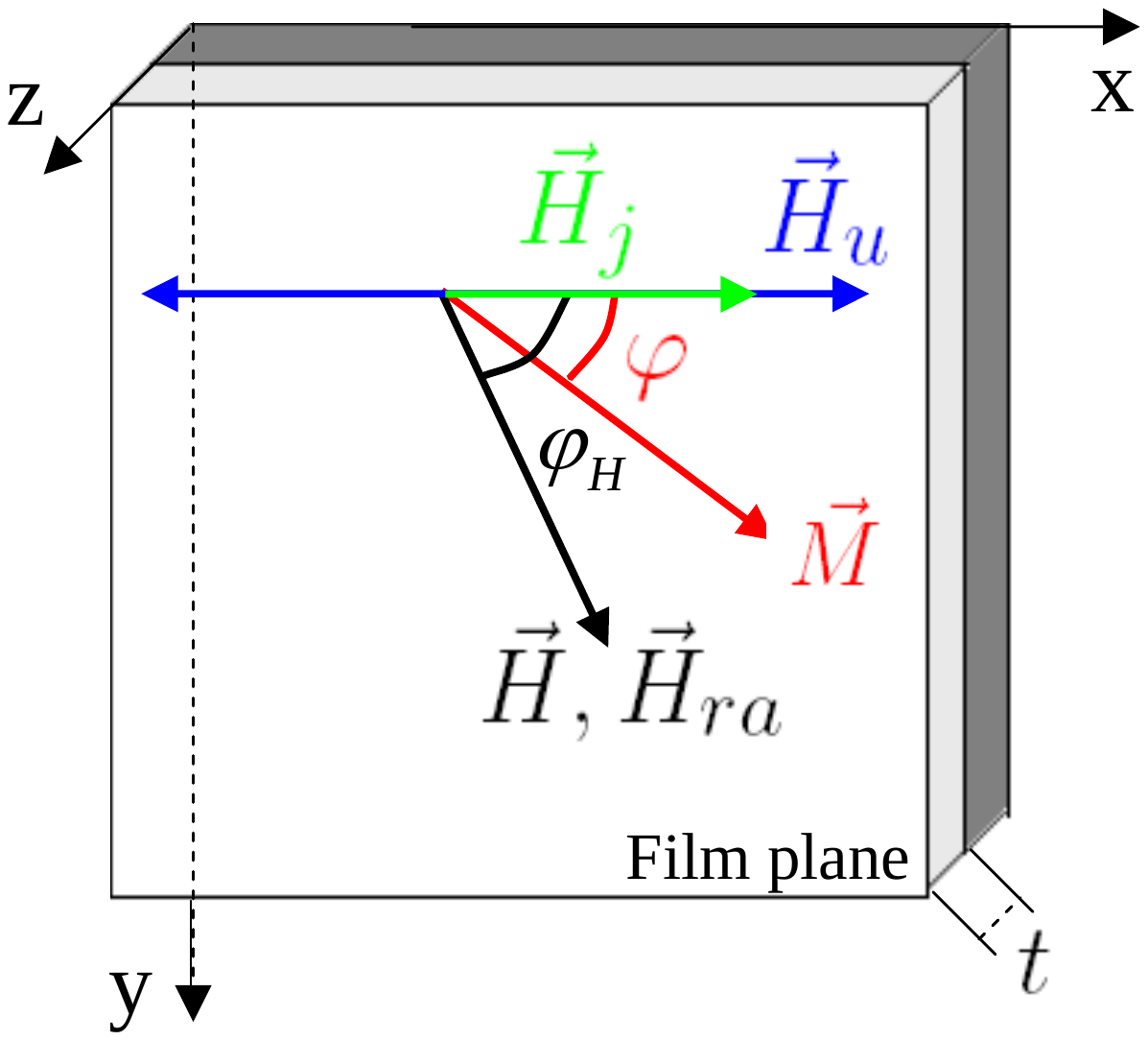}
\caption{Sketch of the coordinate system used with the assumption that $\theta=\pi/2$.
In the experiments presented here, the magnetic field is applied in
the ($xy$) plane. Note that $\vec{H}_{u}$ and $\vec{H}_{j}$ are parallel
one to each other and aligned along $\vec e_x$. $\vec{H}_{ra}$ stands to be
roughly parallel to the in-plane applied magnetic field. }
\end{center}
\end{figure}

The bulk term value was previously found to lie around 10$^{5}$ erg.cm$^{-3}$
\cite{labrune2004}: as discussed in section 4 it is negligible compared to
the experimental determination of $K_{\perp}$. A similar partition
(bulk + surface) could be done for the in-plane anisotropy, but, as
shown in section IV, does not provide for useful conclusions.

We define, as usual, the uniaxial perpendicular and the uniaxial in-plane
anisotropy fields $H_{\perp}$ and $H_{u}$, respectively, as:

\begin{eqnarray}
H_{\perp}=\frac{2K_{\perp}}{M}\,\,\,;\,\,\, H_{u}=\frac{2K_{u}}{M}
\label{eq:4}
\end{eqnarray}

When the permalloy layer is interfaced with an antiferromagnetic layer,
the magnetic energy density includes an additional contribution $\varepsilon_{b}$
\cite{mcmichael}:

\begin{center}
\begin{eqnarray}
\varepsilon_{b}=-\frac{\vec{M}\cdot\vec{v}}{M}j-\frac{\vec{M}\cdot\vec{H}}{MH}j_{ra}\label{eq:5}\end{eqnarray}

\par\end{center}

In equation (5), $j$ and the in-plane vector $\vec{v}$ allow for
defining the unidirectional anisotropy. It is usual to introduce the
exchange bias field $\vec{H}_{j}$:

\begin{center}
\begin{eqnarray}
\vec{H}_{j}=\frac{j}{M}\vec{v}\label{eq:6}\end{eqnarray}

\par\end{center}

The last term in equation (5) stands for a rotatable anisotropy \cite{mcmichael}.
It has not to be taken in consideration for determining the orientation
of the magnetisation at equilibrium but it induces a so-called rotatable
field $\vec{H}_{ra}$ in the equation of motion monitoring the dynamics:

\begin{eqnarray}
\vec{H}_{ra}=\frac{j_{ra}}{MH}\vec{H}
\label{eq:7}
\end{eqnarray}

Generally, $\varepsilon_{b}$ is considered as an effective volume
density arising from a surface contribution depending of two parameters,
$j_S$ and $j_{raS}$: such an assumption provides $j=j_{S}/t$ and $j_{ra}=j_{raS}/t$.

Figure 1 shows the above defined vectors and the used angular notations.
In agreement with our experimental results discussed below we have
taken $\vec{u}$ parallel to $\vec{v}$: this provides simplifications
which are not always satisfied \cite{zighem1,hill1}.

In the present work we specially focused on the uniform magnetic mode,
the frequency $f$ and the linewidth $\Delta f$ of which were measured
by MS-FMR. It results from the Landau-Lifshitz-Gilbert equation of
motion that, for this mode, $f$ and $\Delta f$ are given by \cite{netzelmann}:

\begin{center}
\begin{eqnarray}
\left(\frac{2\pi f}{\gamma}\right)^{2}=\frac{1}{\left(M\sin\theta\right)^{2}}\left(\frac{\partial^{2}\varepsilon}{\partial\theta^{2}}\frac{\partial^{2}\varepsilon}{\partial\varphi^{2}}-\left(\frac{\partial^{2}\varepsilon}{\partial\theta\partial\varphi}\right)^{2}\right)\nonumber \\
\frac{2\pi\bigtriangleup f}{\gamma}=\frac{\alpha}{M}\left(\frac{\partial^{2}\varepsilon}{\partial\theta^{2}}-\frac{1}{\sin^{2}\theta}\frac{\partial^{2}\varepsilon}{\partial\varphi^{2}}\right)
\label{eq:8}\end{eqnarray}

\par\end{center}

In the above expressions $\theta$ and $\varphi$ stands for the polar angles of $\vec M$, 
$\alpha$ is the dimensionless Gilbert coefficient and $\gamma$ is the effective gyromagnetic
factor. For an in-plane applied field ($\Rightarrow\theta=\pi/2$), one obtains:

\begin{eqnarray}
\left(\frac{2\pi f}{\gamma}\right)^{2}=H_{1}\times H_{2}\label{eq:9}
\end{eqnarray}
 with\\
\begin{eqnarray}
H_{1}=((H+H_{ra})\cos(\varphi-\varphi_{H})-H_{u}\cos 2\varphi+H_{j}\cos\varphi)\nonumber\\
\mbox{ and}\nonumber\\
H_{2}=((H+H_{ra})\cos(\varphi-\varphi_{H})+H_{dem.}-H_{\perp}\nonumber\\ +H_{u}\cos^{2}\varphi+ H_{j} \cos \varphi)
\end{eqnarray}

where $H_{dem.}=4\pi M$ is the demagnetizing field. The linewidth
is given by:

\begin{center}
\begin{eqnarray}
\frac{2\pi\bigtriangleup f}{\gamma}=\alpha(2(H\cos(\varphi-\varphi_{H})+H_{u}(3\cos^{2}\varphi-1)\nonumber \\
+H_{j}\cos\varphi)+H_{dem.}-H_{\perp})\label{eq:10}\end{eqnarray}
\par\end{center}

In the case of low applied fields ( $H\ll H_{dem.eff.}=H_{dem.}-H_{\perp}$;
$H_{dem.eff.}$ is the effective demagnetizing field), the linewidth
expression reduces to:\\
\begin{eqnarray}
\left(\frac{2\pi f}{\gamma}\right)^{2}=\alpha H_{dem.eff.}\label{eq:12}\end{eqnarray}

In addition, the so-called DE mode \cite{damon} was studied by BLS. Its
frequency depends on the wave vector $\vec{q}$. It is not given by
an analytic expression but it can be numerically calculated \cite{roussigne95}.
However, for the $qt$ values, small compared to unity, involved
in our Brillouin study, an approximate analytic expression can be
obtained  \cite{zighem2,stamps}. It is given by equation (9), with the following
modified values of and :\\

\begin{eqnarray}
H_{1}=((H+H_{ra})\cos(\varphi-\varphi_{H})+\frac{H_{dem.}}{1+qt/2}-H_{\perp}\nonumber\\ +\frac{2A}{M}q^{2}+H_{u}\cos^{2}\varphi+H_{j}\cos\varphi)\mbox{\,\,\,\,\,\, and} \nonumber\\ 
H_{2}=((H+H_{ra})\cos(\varphi-\varphi_{H})+\frac{H_{dem.}\cos^{2}(\varphi-\varphi_{H})}{1+2/qt} \nonumber\\ +\frac{2A}{M_{s}}q^{2}
 +H_{u}\cos2\varphi+H_{j}\cos\varphi)
\end{eqnarray}

$A$ is the exchange stiffness coefficient. Note that for $q\rightarrow0$,
we retrieve the expression of the uniform mode.

\section{Experimental setups and samples}

\subsection{Experimental setups}

The measurements were performed at room temperature using both MS-FMR
and BLS techniques. In MS-FMR the resonance is probed by sweeping
the frequency of a pumping RF field $h_{rf}$ in presence of a fixed
applied magnetic field. More precisely, the amplitude of the applied
magnetic field is subject to a slight modulation at low frequency
($\sim$ 140 Hz), thus allowing for a synchronous detection system. This
technique gives access to the first derivative of the RF absorption
versus the applied field. This absorption is generally described by
a lorentzian function. The main advantage, compared to conventional
FMR, consists in the availability of resonance studies with various
amplitudes of the applied field and not only with various directions.
Practically, the orders of magnitude of the in-plane anisotropy terms
do not allow for their determination through conventional FMR while
such determinations are easily performed using MS-FMR which is compatible
with the necessarily low values of the applied magnetic fields. However,
with MS-FMR, the rather large lack of spatial homogeneity of the RF
field is a source of distortions of the absorption signal which prevent
for very precise resonance frequency and linewidth evaluations \cite{counil,counil1}. \\

BLS was investigated in order to study the dispersion of the propagating
magnetic mode versus the wave-vector $\vec{q}$ (in the so-called
Damon-Eschbach geometry, which designates geometrical arrangements
characterized by a wave-vector normal to the magnetisation at equilibrium).
The appropriately polarized spectra were studied in retro-scattering
conditions using a $3\times2$ tandem Fabry-Pérot interferometer illuminated
by a single-mode Ar$^{+}$ ion laser at a wavelength of $\Lambda=514.5$
nm with a power of a few hundreds of mW. The sweeping of $q$ was
obtained by varying the angle of incidence of the optical beam ($q=4\pi\sin\psi/\Lambda$
, where $\psi$ is the angle of incidence). In addition, some magnetic
measurements were performed using a vibrating sample magnetometer
(VSM), mainly in view of deriving coercive and bias fields from the
observed magnetisation curves.

\begin{figure}
\begin{center}
\includegraphics[bb=0bp 130bp 595bp 770bp,clip,width=8.5cm]{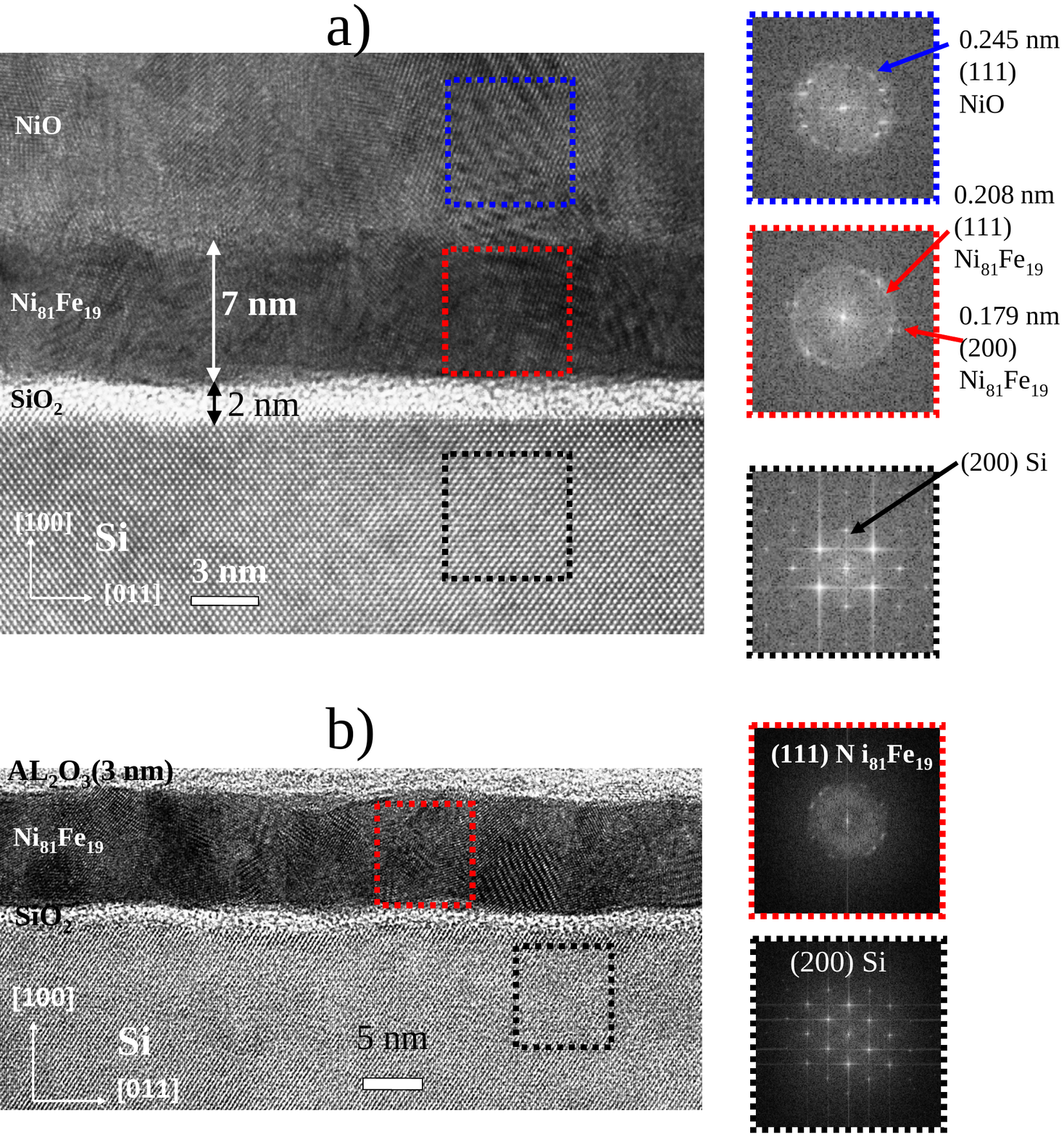}
\caption{HRTEM images of the NiO/Ni$_{81}$Fe$_{19}$ (a) and of the Al$_{2}$O$_{3}$/Ni$_{81}$Fe$_{19}$
(b) interfaces. The colored squares on the right correspond to local Fourier transforms of the respectively  colored areas of the picture.}
\end{center}
\end{figure}

\subsection{Studied Samples}

Two series of samples were elaborated using radio frequency sputtering
on a silicon substrate covered by a thin (2 nm) SiO$_{2}$ layer.
In the first one, the NiO thickness is fixed to 80 nm and the permalloy
thickness $t$ varies from 25 to 5 nm (25, 14, 9, 7.5, 6 and 5 nm).
In the second one, the nickel oxide layer is replaced by a thin film
(3 nm) of alumina (Al$_{2}$O$_{3}$) and the set of permalloy layers
is unchanged : this second series defines a reference in order to
evaluate the interfacial changes related to the antiferromagnetic/ferromagnetic
boundary. High resolution transmission electron micrographs (HRTEM)
were performed (see Figure 2) in view of getting information on the
microstructure of each sub-layer in a given sample. The thicknesses
are found in good agreement with the calculated ones from the deposition
conditions. The NiO and the Ni$_{81}$Fe$_{19}$ films are
well crystallized and composed of small crystallites with sizes of
a few nm$^{3}$). The NiO layers present columns around 5 nm in diameter and 30 nm in length.
Moreover, no preferential texture is favoured, as attested by our electron diffraction studies (not shown here).
This signifies that the NiO interface corresponds to a mixed structure
including compensated and uncompensated magnetic moments. Nevertheless,
the roughness of the interface does not exceed 0.5 nm (see Figure
2). Note also that the two series were studied in the absence of a
preliminary cooling from the Néel temperature under a magnetic field
applied in view of increasing the exchange bias. However, during the
deposition a small residual in-plane magnetic field of 5 Oe is applied
along $\vec e_x$, as above mentioned: it favours the interfacial
exchange coupling.

\section{Results and discussion}

\subsection{Unbiased Al$_{2}$O$_{3}$/permalloy films}

For this reference series the contribution $\varepsilon_{b}$, expressed
in equation (5), vanishes. The frequency and the linewidth of the uniform mode are expected
to only depend on the gyromagnetic factor $\gamma$, of the magnetisation
$M$, of the uniaxial perpendicular anisotropy ($K_{\perp}$), of
the uniaxial in-plane anisotropy ($K_{u}$) and of the damping ($\alpha$)
coefficients (and, indeed, of the sample thickness $t$). In principle
the study of their variations versus the amplitude and the direction
of an applied magnetic field allows for the evaluation of these magnetic
parameters. However, due to the limitations of the available precision,
the MS-FMR study is mainly efficient to give access to $H_{dem.eff.}$,
$H_{u}$ and $\alpha$, assuming a given value of $\gamma$ (we took
1.844$\times10^{7}$ s$^{-1}$.Oe$^{-1}$ in agreement with the expected
value (2.1) of the effective $g$ factor ($\gamma=g\times8.794\times10^{6}$)).\\

\begin{figure}[!h]
\begin{center}
\includegraphics[bb=25bp 25bp 360bp 595bp,clip,width=8.5cm]{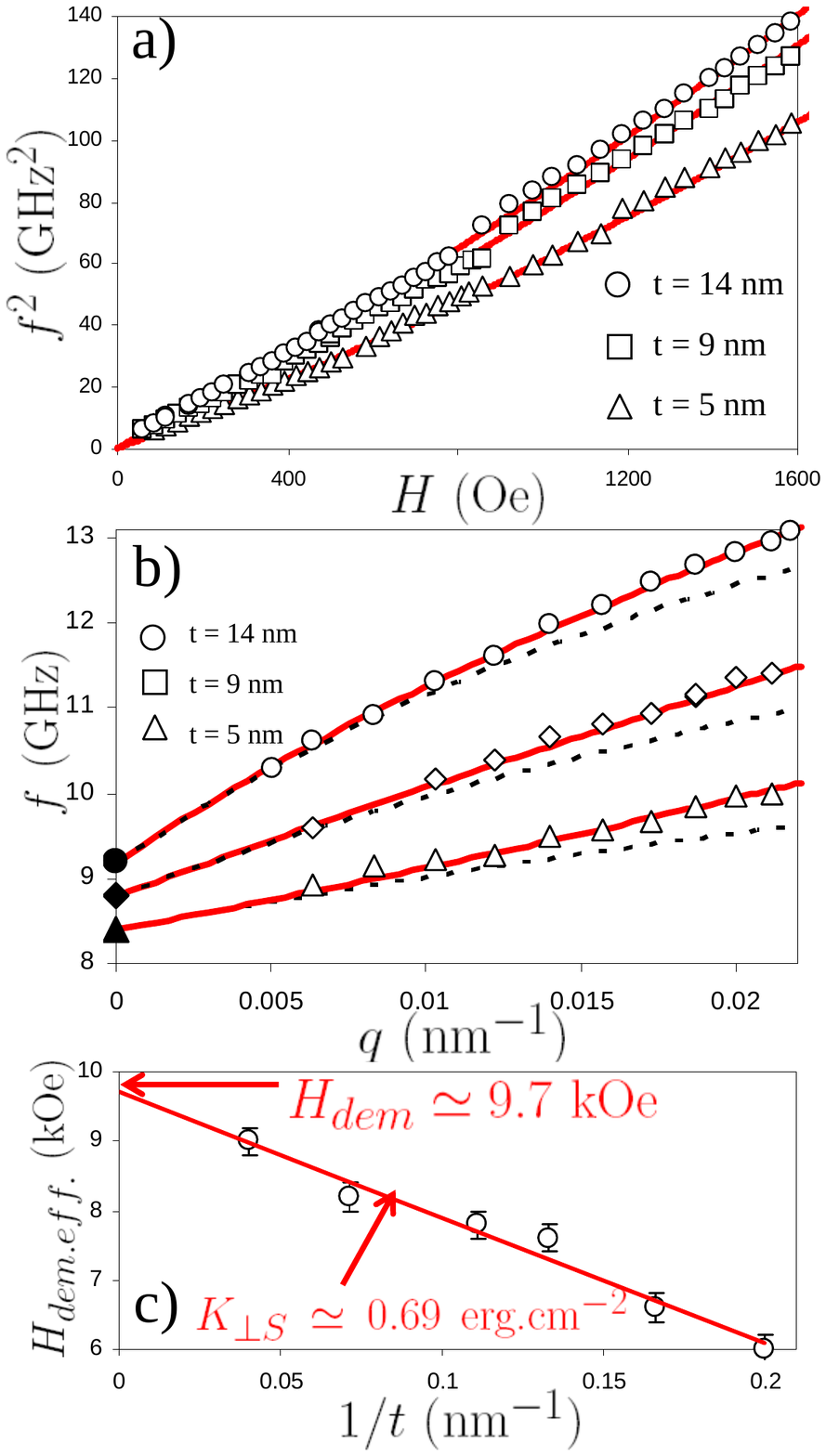}
\caption{a) Squared resonance frequency versus the applied magnetic field along
the easy axis ($\vec e_x$ direction). b) $q$-dependence of the DE mode frequency for three characteristic
thicknesses. Full lines correspond to the adjustments obtained by
using equation 10 and 13. We have also reported the frequency of
the uniform mode ($q=0$) measured by MS-FMR. The dashed lines correspond
to the fit neglecting the exchange stiffness ($A=0$). c) Deduced
$H{}_{dem.eff.}$ values versus the inverse of the permalloy thickness: the found anisotropy constant
value is $K_{\perp S}\simeq0.69$ erg.cm$^{-2}$. }
\end{center}
\end{figure}

The variations of the frequency versus the amplitude (see Figure 3)
and the direction (see Figure 5) of an in-plane applied field were
studied. It results from the analysis of the data that the surface
coefficient $2K_{\perp S}/M$ contributing to the perpendicular anisotropy
does not depend on the studied sample. From its observed linear variation
versus ($1/t$) we deduce a common value of the effective demagnetizing
field at large thicknesses, equal to 9700 Oe (see Figure 3c). This value is very close
to the reported one in bulk permalloy showing the same composition
\cite{lykken,rantschler}. As mentioned in section 2, $2K_{\perp B}/M$ is not
expected to exceed a few hundred Oe. With a rather good approximation,
one can neglect this bulk anisotropy and conclude that, in the studied
interval of thicknesses, all the samples show a saturation magnetisation
practically equal to the measured one in the bulk material and are
characterized by the same surface energy $K_{\perp S}=$0.69 erg.cm$^{-2}$.
Considering now the in-plane anisotropy, it is found very small, thus
providing anisotropy fields of a few Gauss, with an easy axis parallel
to the direction of the field induced during the elaboration process.
In addition, the Gilbert damping model provides a satisfactory agreement
of the observed linewidths: as expected from this model the deduced
value of $\alpha$ (see Figure 4b) does not depend on the applied field. It varies
from one to another sample but always lies in the {[}0.005; 0.009{]}
interval.

\begin{figure}[!]
\begin{center}
\includegraphics[bb=20bp 30bp 470bp 600bp,clip,width=8cm]{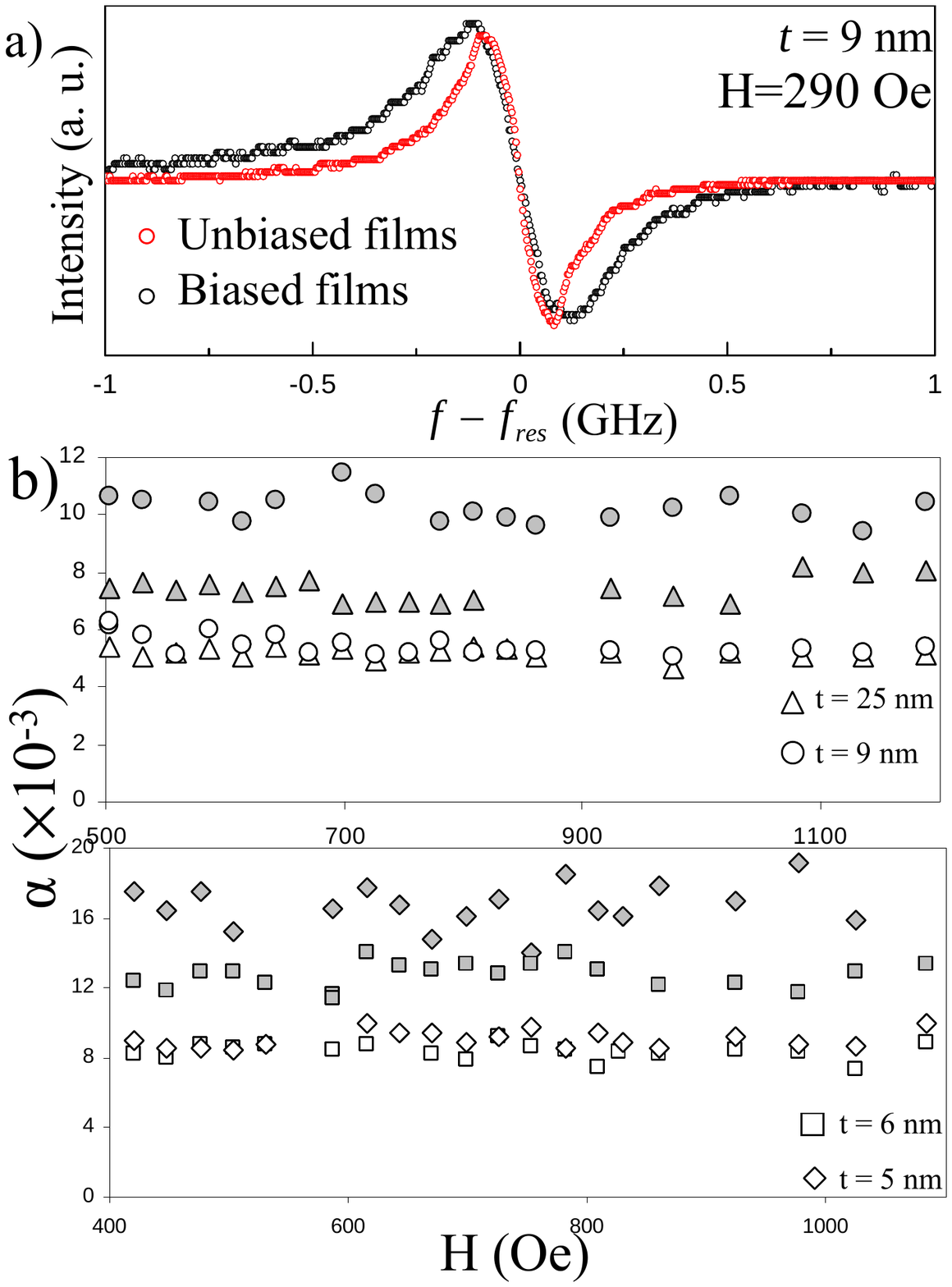}
\caption{a) Typical spectra obtained from the two series of samples.
Note the linewidth broadening obtained in presence of the nickel oxide
layer. These examples correspond to spectra obtained from the samples
of $t=9$ nm under an applied field of 290 Oe. For clarity, they are
centered around 0 (shifted by their respective resonance frequency).
b) Damping parameter $\alpha$ versus the applied magnetic field.
Open symbols represent unbiased films while filled symbols represent
the results obtained from biased films.}
\end{center}
\end{figure}

The anisotropy and magnetisation values calculated from the MS-FMR
data provide a good fit of the Brillouin spectra: as written above,
the $q$ dependence of the frequency is deduced from its variation
versus the angle of incidence of the illuminating beam. When its direction
is nearly normal to the film this frequency practically does not differ
from the obtained one for the uniform mode, as illustrated on Figure
3b. For large $q$ values the contribution of the exchange stiffness
constant $A$ is significant and has to be taken in account (see equation
(13) and Figure 3b). For all the studied samples our measured value
of $A$ is equal to $1\times10^{-6}$ erg.cm$^{-1}$. However, this
determination is rather approximate: to improve the precision it would
be useful to observe the stationary magnetic modes \cite{belmegue} which, due to limitations
in the available amplitude of the applied field, were not accessible
in the present study. The complete set of the magnetic parameters
deduced for the studied series of unbiased Al$_{2}$O$_{3}$/permalloy
films is given in Table 1.

\begin{center}
\begin{tabular}{cccc}
\hline\hline
$t$   & $H_{dem.eff.}$  &    $H_u$     &    $\alpha$\\
 (nm)  &  (kOe)  &  (Oe)    &    $\times10^{-3}$\\
\hline
\hline
25  & 9  & $\sim$ 5 &5 \tabularnewline
14  & 8.2  & $\sim$ 5&6 \tabularnewline
9  & 7.8  & $\sim$ 5&5 \tabularnewline
7.5  & 7.6  &$\sim$ 5 &7\tabularnewline
6  & 6.6  & $\sim$ 5&9 \tabularnewline
5  & 6  & $\sim$ 5&8 \tabularnewline
\hline
\hline
\multicolumn{4}{c}{$H_{dem.}=9.7$ kOe;
 $K_{\perp S}\simeq0.69$ erg.cm$^{-2}$; }\\
 \multicolumn{4}{c}{$A=1\times10^{-6}$ erg.cm$^{-1}$ }\\
\hline
\hline
\multicolumn{4}{c}{Table I: Unbiased Al$_{2}$O$_{3}$/permalloy films}\tabularnewline
\end{tabular}
\par\end{center}

\subsection{Biased NiO/permalloy films}

The magnetisation curves, recorded with an in-plane applied magnetic
field, clearly reveal an easy anisotropy axis and a bias field along the above
mentioned residual magnetic field applied in $\vec e_x$ direction during
the deposition. Figure 5 shows examples of hysteresis graphs obtained
along easy and hard directions. It also presents the variation versus
$\varphi_{H}$ of the remanent magnetisation after suppressing a saturating
magnetic field inclined of $\varphi_{H}$ from $\vec e_x$. The results can
be interpreted in terms of a coherent uniform rotation model depending
of an energy density containing both contributions of a coercive field
($H_{c}$) and of a bias field ($H_{d}$). The obtained values of
($H_{c}$) and of ($H_{d}$) are given in Table II. The MS-FMR measurements
allow for evaluating the parameters enumerated in the previous section
and, in addition, the bias and the rotatable fields appearing in such
biased structures. \\

\begin{figure}[!]
\begin{center}
\includegraphics[bb=25bp 135bp 500bp 600bp,clip,width=8.5cm]{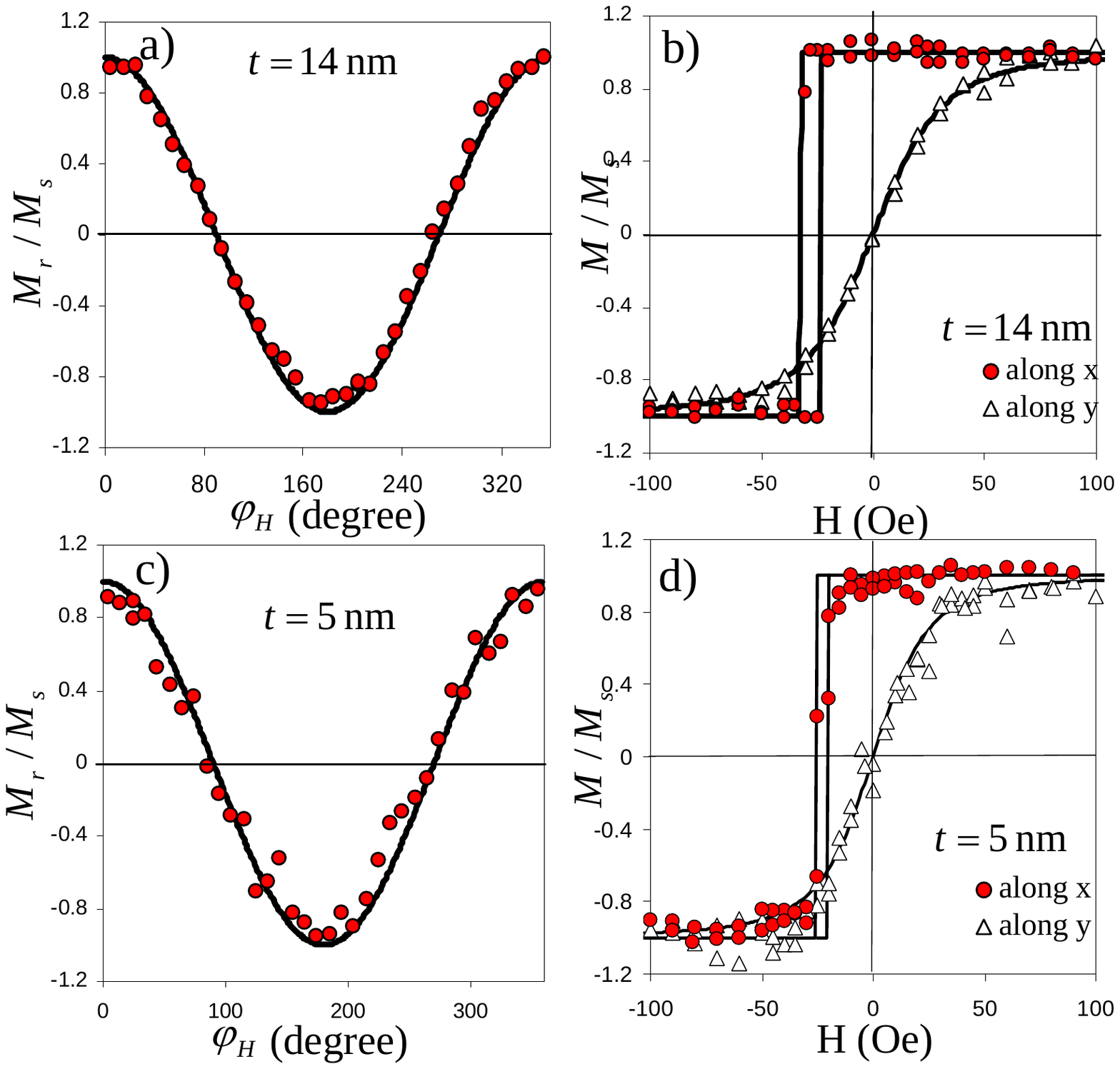}
\caption{a) c) Angular dependence of the remanent magnetisation for two typical
thicknesses (14 and 5 nm). Magnetisation curves measured at $\varphi_{H}=0^{\circ}$(along
$\vec e_x$) and $\varphi_{H}=90^{\circ}$ (along $\vec e_y$). The solid lines corresponds to the
best fits obtained from the magnetic energy density presented in equation
1.}
\end{center}
\end{figure}

The first important result concerns the effective
demagnetizing field: it practically does not differ from the measured
one in the unbiased films described in the previous subsection. We
then conclude to a perpendicular anisotropy coefficient mainly monitored
by a surface term $K_{\perp S}$. We derive: $K_{\perp S}=$0.65 erg.cm$^{-2}$,
nearly the same value as in the unbiased samples (0.69 erg.cm$^{-2}$).
Here again, the saturation magnetisation does not appreciably differ
from the reported value in the bulk \cite{lykken,rantschler}. The in-plane anisotropy parameters
involved in these structures, $H_{u}$, $H_{j}$ and $H_{ra}$, are
presented in Table II. Typical angular variations of the frequency
versus the direction of the in-plane applied field, compared to the
observed ones in unbiased samples of the same thickness, are shown
on Figure 6. The uniaxial anisotropy field $H_{u}$ does not overpass
a few Oe, as it is the case in the unbiased samples. Notice that it
does not significantly differ from the above discussed coercive field,
as expected in the frame of a coherent uniform rotation model. A bias
exchange field is observed in all the films: here again, it does not
much differ from the static bias field $H_{d}$ deduced from the hysteresis
loops. In both cases there is no clear correlation between the in-plane
anisotropy values and the thickness of the studied film. Table II
also gives the Gilbert damping coefficients  derived from the analysis
of the linewidths of the studied resonances: $\alpha$ is larger than
in the unbiased layers and increases when the thickness decreases.
This behaviour was pointed out in several publications \cite{rezende1,mcmichael,zighem1}.
A theoretical model introduced by Arias et al. \cite{mills} was adapted
by Rezende et al. \cite{rezende2} in order to explain this broadening. In
this model, it results from the lack of homogeneity of the interfacial
exchange coupling between the ferromagnetic and antiferromagnetic
layers and varies as $t{}^{-2}$ . The magnetic parameters derived
from our MS-FMR measurements are in agreement with the BLS results.
The $q$ dependence of the observed frequency is well accounted assuming
an unchanged value of the exchange stiffness $A$ ($1\times10^{-6}$
erg.cm$^{-1}$). However, due to the rather small in-plane anisotropy
terms and to the limited available instrumental precision, their quantitative
derivation through our BLS study is not very efficient.\\
\\
\\

\begin{center}
\begin{tabular}{cccccccc}
\hline
$t$  & $H{}_{dem.eff.}$  & $H_{u}$  & $H_{c}$  & $H_{j}$  &  $H_d$  & $H_{ra}$ &$\alpha_{app}$($\times10^{-3}$) \tabularnewline
(nm)  & (kOe)  & (Oe)  & (Oe)  & (Oe)   & (Oe) & (Oe) & ----- \tabularnewline
\hline
\hline
25  & 9    & 6  & 5 &10   & 10 & 0    & 7    \tabularnewline
14  & 8.4  & 8  & 5 & 20  & 23 & 15  & 9    \tabularnewline
9   & 7.6  & 4  & 4 & 18  & 30 & 8   & 10   \tabularnewline
7.5 & 7.3  & 3  & 8 & 32  & 41 & 18  & 14   \tabularnewline
6   & 7    & 3  & 5 & 27  & 24 & 15  & 12   \tabularnewline
5   & 5.1  & 2  & 3 & 21  & 28 & 25  & 16   \tabularnewline
\hline
\hline
\multicolumn{8}{c}{ $H_{dem.}=9.7$ kOe; $K_{\perp s}\simeq0.65$ erg.cm$^{-2}$ }  \tabularnewline
\multicolumn{8}{c}{  $A=1\times10^{-6}$ erg.cm$^{-1}$}  \tabularnewline
\hline
\hline
\multicolumn{8}{c}{Table II: Biased NiO/permalloy films} \tabularnewline
\end{tabular}
\par\end{center}

\begin{figure}[!]
\begin{center}
\includegraphics[bb=30bp 65bp 480bp 590bp,clip,width=8.5cm]{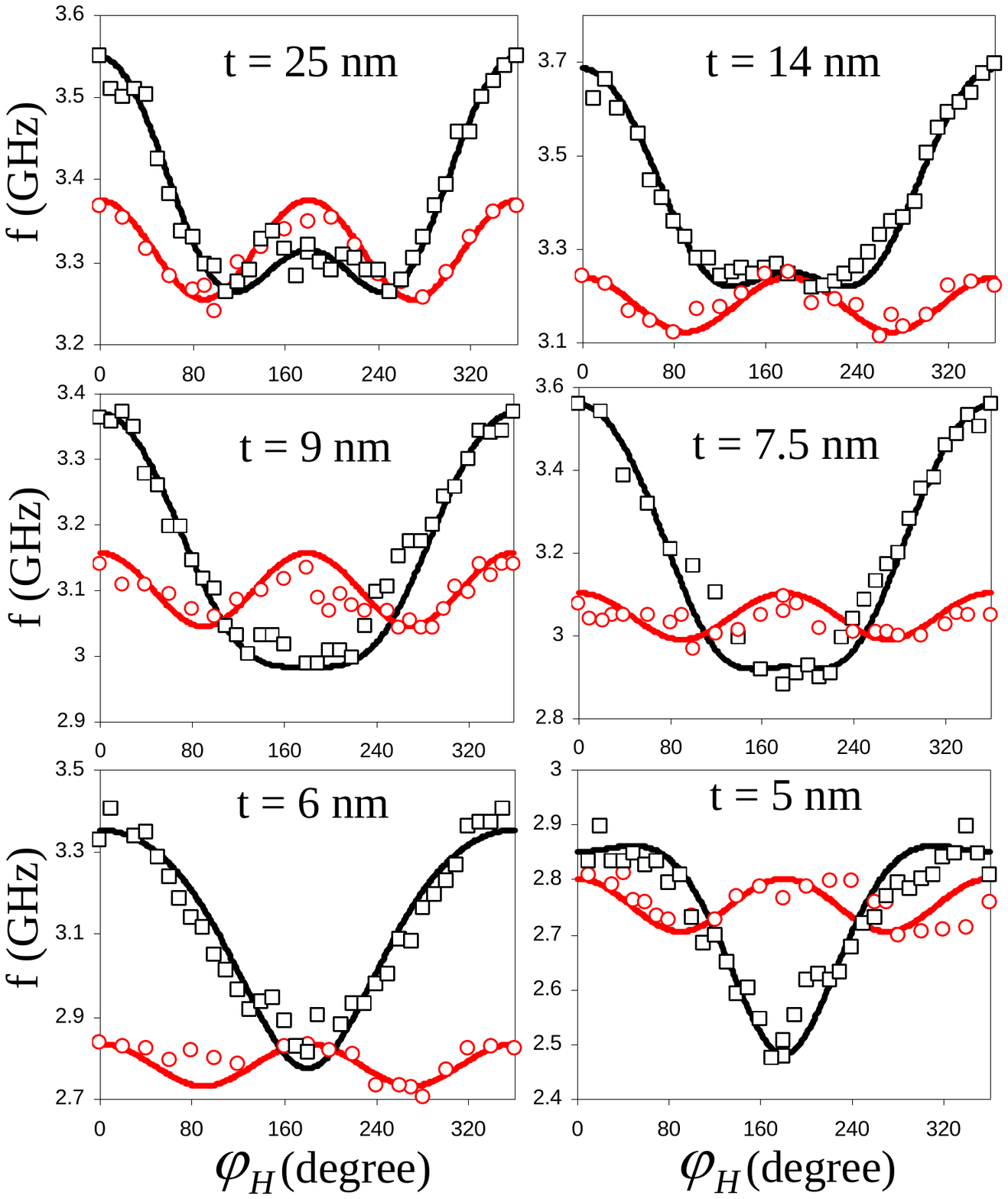}
\caption{Angular dependence of the uniform mode frequency. A small in-plane
applied magnetic field (140 Oe) was fixed to ensure the saturation
magnetisation. The black squares correspond to the results obtained
from the biased samples and the circles are the results obtained from the
unbiased samples. The solid lines correspond to the best fits as calculated
using equation 10. The different parameters used are presented
on Table I and II.}
\end{center}
\end{figure}

\section{Conclusion}

Our comparative study of two series of permalloy ferromagnetic (F)
layers covering a thickness interval extending from 5 to 25 nm, grown
by RF sputtering, respectively on a non-magnetic oxide substrate (Al$_{2}$O$_{3}$)
and on an antiferromagnetic (AF) one (NiO), revealed expected differences
originating from the exchange F/AF interfacial interaction but also
some surprising similarities. In both cases the interface, observed
by HRTEM, shows a very small roughness (below 0.5 nm). The dynamic magnetic
properties put in evidence an uniaxial perpendicular anisotropy mainly
originating from a contribution of the surface density of energy which is practically
independent of the nature of the interface (Al$_{2}$O$_{3}$/permalloy
or NiO/permalloy). The saturation magnetisation does not vary versus
the thickness and is close to the expected one in bulk permalloy
showing the same composition. The in-plane anisotropy terms derived
from MS-FMR data are in reasonable agreement with the deduced ones
from conventional VSM magnetometry. The in-plane anisotropy field
does not exceed a few Oe and is close to the coercive field ; its
value is not clearly correlated neither to the thickness nor to the
nature of the interface. The bias exchange is only observed in presence
of a AF/F interface, as usual.


\begin{thebibliography}{30}
\bibitem{neel} L. Néel, J. Phys. Radium \textbf{15}, 225 (1954)

\bibitem{bruno} C. Chappert and P. Bruno, J. Appl. Phys \textbf{64},
5736 (1988)

\bibitem{roussigne95} Y. Roussigné, F. Ganot, C. Dugautier and D.
Renard, Phys. Rev. B \textbf{52}, 350 (1995)

\bibitem{meiklejohn} W. H. Meiklejohn and C. P. Bean, Phys. Rev.
\textbf{102}, 1413 (1956)

\bibitem{nogues} J. Nogues, I. K. Schuller, J. Magn. Magn. Mater.
\textbf{192}, 203 (1999)

\bibitem{binek} C. Binek, Phys. Rev. B \textbf{70,} 014421 (2004)

\bibitem{mcmichael} R. D. McMichael, M. D. Stiles, P. J. Chen and
W. F. Egelhoff, Phys. Rev. B \textbf{58}, 8605 (1998)

\bibitem{nogues2005}J. Nogués, J. Sort, V. Langlais, V. Skumryev,
S. Suriñach, J.S. Muñoz, M.D. Baró, Physics Reports \textbf{422},
65 (2005).

\bibitem{berkowitz1999}A.E. Berkowitz and K. Takano, J. Magn. Magn.
Mater. \textbf{200}, 552 (1999).

\bibitem{stamps2000}R. L. Stamps, J. Phys. D: Appl. Phys. \textbf{33,}
R247 (2000)


\bibitem{stoecklein} W. Stoecklein, S. S. P. Parkin and J. C. Scott,
Phys. Rev. B \textbf{38}, 6847 (1988)


\bibitem{geshev} J. Geshev, L. G. Pereira, J. E. Schmidt, L. C. C.
M. Nagamine, E. B. Saitovitch and F. Pelegrini, Phys. Rev. B \textbf{67},
132401 (1998)


\bibitem{blachowicz} T. Blachowicz, J. Appl. Phys. \textbf{102},
043901 (2007)

\bibitem{wee} L. Wee, R. L. Stamps, L. Malkinski and Z. Celinski,
Phys. Rev. B \textbf{69} 043901 (2007)

\bibitem{rezende1} S. M. Rezende, M. A. Lucena, A. Azevedo, F. M.
de Aguiar, J. R. Fermin and S. S. P. Parkin, J. Appl. Phys. \textbf{93}
7714 (2003)

\bibitem{zighem1}F. Zighem, Y. Roussigné, S.-M. Chérif and P. Moch,
J. Phys. Cond. Matter, \textbf{20}, 125201 (2008)

\bibitem{labrune2004} J. Ben Youssef, N. Vukadinovic, D. Billet and
M. Labrune Phys. Rev. B \textbf{69}, 174402 (2004)


\bibitem{hill1} M. O. Liedke, B. Liedke, A. Keller, B. Hillebrands,
A. Mücklich, S. Facsko and J. Fassbender, Phys. Rev. B, \textbf{75},
220407 (2007)

\bibitem{netzelmann} U. Netzelmann, J. Appl. Phys. \textbf{68}, 1800
(1990)

\bibitem{damon} W. Damon and J. R. Eshbach J. Phys. Chem. Solids,
\textbf{19}, 308 (1961)

\bibitem{zighem2}F. Zighem, Y. Roussigné, S.-M. Chérif and P. Moch,
J. Phys. Cond. Matter, \textbf{19}, 176220 (2007)

\bibitem{stamps} R. L. Stamps Phys. Rev. B, \textbf{49}, 339 (1994)



\bibitem{counil} G. Counil, J.-V. Kim, T. Devolder,  P. Crozat, C. Chappert, K. Shigeto and Y. Otani
J. Appl. Phys. \textbf{95}, 5646 (2004)

\bibitem{counil1} G. Counil, J.-V. Kim, T. Devolder, C. Chappert and A. Cebollada
J. Appl. Phys. \textbf{98}, 023901 (2005)


\bibitem{lykken} G. I. Lykken, W. L. Harman and E. N. Mitchell, J.
Appl. Phys. \textbf{37}, 3353 (1966)

\bibitem{rantschler}J. O. Rantschler, P. J. Chen, A. S. Arrott, R. D. McMichael, W. F. Egelhoff, Jr. and B. B. Maranville,
J. Appl. Phys. \textbf{97}, 10J113 (2005)

\bibitem{belmegue} M. Belmeguenai, F. Zighem, Y. Roussigné, S.-M.
Chérif, P. Moch, K. Westerholt, G. Woltersdorf and G. Bayreuther,
Phys. Rev. B, \textbf{79}, 024419 (2009)


\bibitem{mills} R. Arias and D. L. Mills, Phys. Rev. B \textbf{60},
7395 (1999)

\bibitem{rezende2} S. M. Rezende, A. Azevedo, M. A. Lucena and F.
M. de Aguiar , Phys. Rev. B \textbf{63}, 214418 (2001)








\end{thebibliography}
\end{document}